\documentclass[]{aa}
\usepackage{natbib}
\usepackage{graphicx}
\usepackage{wasysym}
\usepackage{txfonts}
\usepackage{color}
\def\rmdv{\mbox{\rm dv}}
\def\fH2{\mbox{f$_\HH$}}
\def\EBV{\mbox{E(B-V)}}

\def\nH2{\mbox{${\rm n}_\HH}$}

\def\pccc{~{\rm cm}^{-3}} 
 
\def\pcc{~{\rm cm}^{-2}}

\def\Tsub#1 {\mbox{${\rm T}_{\rm #1}$}}
\def\TK  {\Tsub K }

\def\Tex {\Tsub ex }

 \def\arcmin{\mbox{$^{\prime}$}}

\def\degr{$^{\rm o}$}
\def\p{\mbox{$^+$}}

\def\XCOa{\mbox{X$^0_{\rm CO}$}}

\def\XCO{\mbox{X$_{\rm CO}$}}
\def\ICO{\mbox{$\Upsilon_{\rm CO}$}}
\def\Ihcop{\mbox{$\Upsilon_\hcop$}}

\def\cch{\mbox{C$_2$H}}

\def\h13cop{\mbox{{H$^{13}$CO\p}}}

\def\C3H{\mbox{C$_3$H}}
\def\c3h2{\mbox{C$_3$H$_2$}}
\def\cc3h2{\mbox{{\it c}-C$_3$H$_2$}}
 
 \def\R0{R$_0$}
\def\G0{\mbox{G$_0$}}

\def\ddeg{{}^\circ\kern-.1em}

\def \kms{\mbox{km\,s$^{-1}$}}

\def\E#1 {$10^{#1}$}
\def\E#1 {E{#1}}
\def\P#1,{$\nH2\TK~=~#1\times~10^4\pccc$~K}
\def\ec#1,#2,#3,{#1\,(#2)\E{#3}}

\def\H3{\mbox{H$_3$}}

\def\RH2{\mbox{R$_{\rm G}$}}
\def\g13{\mbox{g$_{13}$}} 

\def\cc3h{\mbox{{\it c}-\C3H}}
\def\lc3h{\mbox{{\it l}-\C3H}}

\newcommand{\emm}[1]{\ensuremath{#1}}   
\newcommand{\emr}[1]{\emm{\mathrm{#1}}} 

\newcommand{\hcop}{\emr{HCO^+}} 
 
\newcommand{\HH}{\emr{H_2}}
\newcommand{\cotw}{\emr{^{12}CO}}
\renewcommand{\coth}{\emr{^{13}CO}}

\newcommand{\N}[1]{\emr{N_{#1}}}

\newcommand{\W}[1]{\emm{{\rm W}_\emr{#1}}}
\newcommand{\WCO}{\W{CO}}

\def\tJ{\tau_{J,J+1}}
\def\t01{\tau_{0,1}}
\def\taup{$\tau_{353}$}

\title{Standing in the shadow of dark gas: ALMA observations of absorption from dark CO in the molecular DNM of Chamaeleon  }

\author{ H. Liszt\inst{1} and M. Gerin\inst{2} and I. Grenier\inst{3}}

\institute{
     National Radio Astronomy Observatory,
           520 Edgemont Road,
           Charlottesville, VA,
           USA 22903 
      \email{hliszt@nrao.edu}
\and
LERMA, Observatoire de Paris,  PSL Research University, CNRS
Sorbonne Universit\'es, UPMC Univ. Paris 06, Ecole Normale
  Sup\'erieure, F-75005 Paris, France
\email{maryvonne.gerin@ens.fr}
\and
AIM, CEA-IRFU/CNRS/Universit\'e Paris Diderot,
D\'epartement d'Astrophysique, CEA,
Saclay, 91191, Gif-sur-Yvette, France
\email{isabelle.grenier@cea.fr}
}

\begin{document}
\date{received \today}%
\offprints{H. S. Liszt}%
\mail{hliszt@nrao.edu}%
%
\abstract
{We previously detected 89.2 GHz J=1-0 \hcop\ absorption in 12 directions 
lacking detected CO emission in the outskirts of the Chamaeleon cloud complex
and toward one sightline with integrated CO emission \WCO = 2.4 K-\kms.  
Eight sightlines had a much larger mean column density of dark 
neutral medium (DNM) - gas not represented in H I or CO emission - and were 
found to have much higher mean molecular column density.  The five other 
sightlines had little or no  DNM and were found to have much smaller 
but still detectable N(\hcop). }
{To determine the CO column density along previously-observed
Chamaeleon sightlines and to determine why CO emission
was not detected in directions where molecular gas is present. }
{We took \cotw\ J=1-0 absorption profiles toward five sightlines
having higher DNM and \hcop\ column densities and one sightline with smaller 
N(DNM) and N(\hcop).  We converted the integrated \hcop\ optical depths to 
N(\HH) in the weak-excitation limit using N(\hcop)/N(\HH) $= 3\times 10^{-9}$ 
and converted the integrated 
CO optical depths \ICO\ to CO column density using the relationship 
N(CO) $= 1.861\times 10^{15}\pcc \ICO^{1.131}$
found along comparable lines of sight that were previously studied in
J=1-0 and J=2-1 CO absorption and emission.}
{CO absorption was detected along the five sightlines in the higher-DNM group, 
 with CO column densities $4\times 10^{13} \pcc \la $ N(CO) $\la 10^{15}\pcc$ that 
 are generally below the detectability limit of CO emission surveys. }
 {In the outskirts of the Chamaeleon complex, the presence of molecular DNM resulted
 primarily from small CO column densities at the onset of CO formation 
 around the HI/\HH\ transition in diffuse molecular gas.  CO relative abundances
  N(CO)/\HH\ $\la 2\times 10^{-6}$ in the outskirts of Chamaeleon are comparable to 
 those seen in UV absorption toward early-type stars, including in Chamaeleon.}


\authorrunning{Liszt, Grenier, Gerin} \titlerunning{Standing in the shadow of the dark CO}

\maketitle{}

%

\section{Introduction}

The total gas column densities that are jointly traced by dust and by cosmic-ray 
interactions exhibit large excesses over those inferred from HI and CO line 
emission in galactic molecular cloud complexes \citep{GreCas+05}, including
in Chamaeleon \citep{Pla15Cham}.  In a recent paper  \citep{LisGer+18} (Paper I) 
we sought evidence of a hidden molecular gas reservoir in Chamaeleon. We presented 
observations of absorption from $\lambda3$ mm J=1-0 lines \hcop, HCN, and \cch\ 
toward thirteen compact extragalactic continuum sources seen against the outskirts 
of the Chamaeleon cloud complex.  We detected \hcop\ absorption in all directions 
even though CO emission was firmly detected in only one of these.  

Taking an abundance ratio N(\hcop)/N(\HH) $ = 3\times 10^{-9}$ we compared the 
inferred column densities of \HH\ to those of the dark neutral medium (DNM), the 
gas inferred to be present from maps of sub-mm dust opacity and gamma-ray emissivity 
\citep{Pla15Cham,RemGre+18} but not apparently represented in $\lambda$21 cm HI or 
$\lambda$2.6mm CO line emission.  We found that gas in the outskirts of Chamaeleon 
was mostly atomic (H I rather than \HH) but that the DNM was mostly molecular, resulting 
from failure of the integrated CO J=1-0 emission \WCO\ to represent the true amount of 
molecular hydrogen when using a standard CO-\HH\ conversion factor 
N(\HH)/\WCO\ = $2\times 10^{20}\pcc$ (K-\kms)$^{-1}$.  Saturation of 
the HI emission profile might have contributed significantly to N(DNM) in three 
directions but did not create the need for the DNM.

Here we ask why CO emission did not represent the molecular gas component
adequately: perhaps as the result of inadequate rotational excitation of the CO, 
or perhaps as the result of there being only very small CO column densities at 
the onset of CO formation around the H I - \HH\ transition.  In the present work 
we observed $\lambda$2.6 mm J=1-0 CO absorption toward six of the thirteen
sources observed in our earlier paper, five in the group of eight directions showing
more DNM and \HH, and (because it was convenient) one direction in the group of five 
showing little or no DNM and much smaller but still non-zero N(\hcop).  The integrated 
CO optical depths are converted to CO column density using the tight relationship
defined along comparable lines of sight previously studied in CO J=1-0 and J=2-1
absorption and emission by \cite{LisLuc98}.

The organization of this paper is as follows.  In Sect. 2 we discuss the  new and 
existing observational material that is presented here and give the background of 
the integrated optical depth - column density conversion for the CO J=1-0 absorption
profiles.  Sect. 3 presents the results of this work and, in Sect. 4, drawing on 
earlier studies of the excitation of CO, we show that the failure of CO emission to 
represent N(\HH) is due to small CO column densities N(CO) and relative abundances 
N(CO)/N(\HH) that are familiar from studies of UV absorption at comparable reddening 
\EBV.  Sect. 5 slightly extends the DNM analysis of Paper I using direct UV absorption
measurements of N(\HH) and N(CO) toward a handful of stars seen toward Chamaeleon
 and Sect. 6 is a summary.
 
\section{Observations, data reduction and optical depth - column density conversion}

\subsection{Conventions}

In this work, N(H) is the column density of H-nuclei in neutral atomic and 
molecular form, N(H) = N(H I)+2N(\HH).  We denote the observed, integrated 
J=1-0 optical depths of the J=1-0 CO and \hcop\ transitions as 
\ICO\ $= \int \t01({\rm CO}) ~\rmdv$ and \Ihcop\ $= \int \t01(\hcop) ~\rmdv$.   
Velocities presented with the spectra are measured with respect to the 
kinematic definition of the Local Standard of Rest.

\begin{figure*}
\includegraphics[height=14cm]{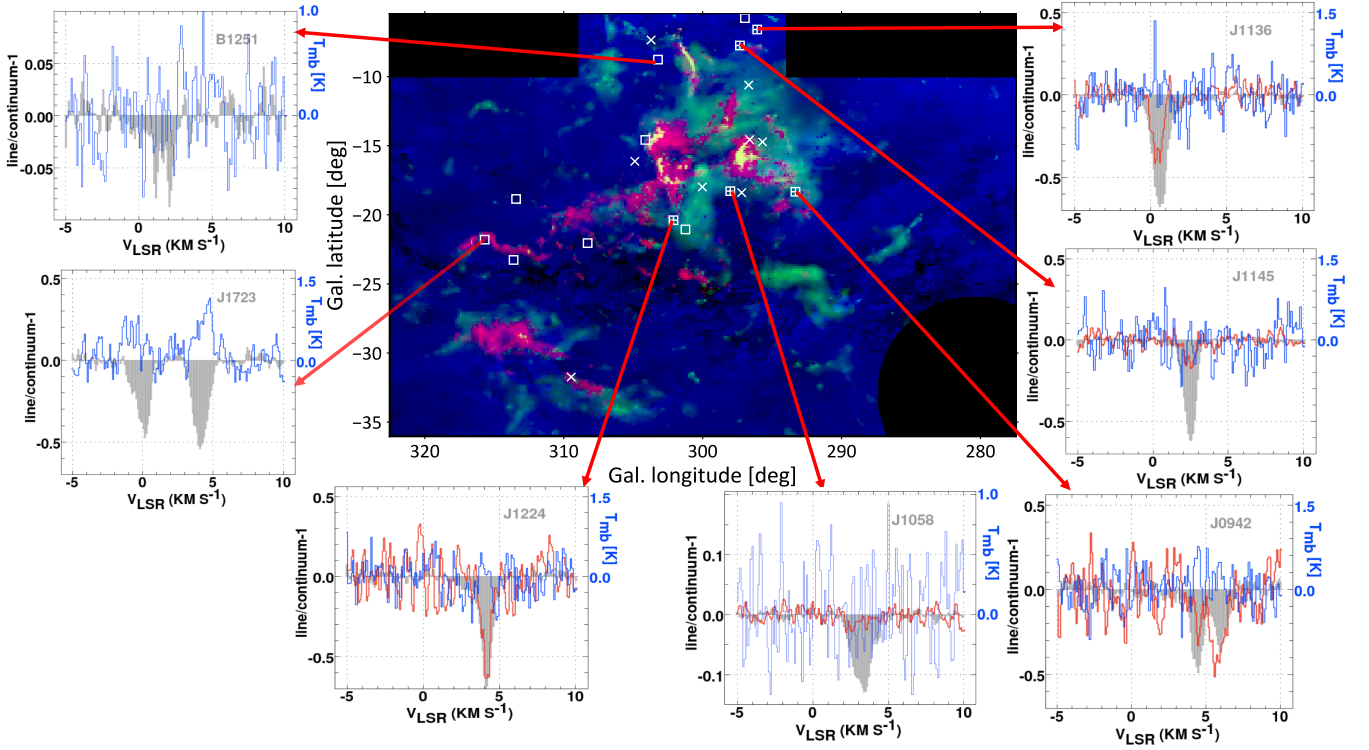}
  \caption[]{ALMA results in the Chamaeleon complex.  Center: RGB image
with N(HI) in blue, N(DNM) in green and Nanten CO emission brightness in red. The 
locations of the 13 ALMA sources studied in Paper 1 are marked as squares, hatched
for the 5 directions where CO absorption was detected in this work. The positions 
of 8 stars with measured values of N(\HH) and N(CO) in UV
absorption (Table 3) are marked by x's. Shown 
around the outside are \hcop\ (grey, shaded) and CO (red) absorption
and CO emission profiles (blue) from NANTEN on a different vertical scale as
noted at right in each outer panel.  CO absorption was not observed toward J1723
and was sought but not detected or plotted toward B1251.}
\end{figure*}

\begin{figure*}
\includegraphics[height=11cm]{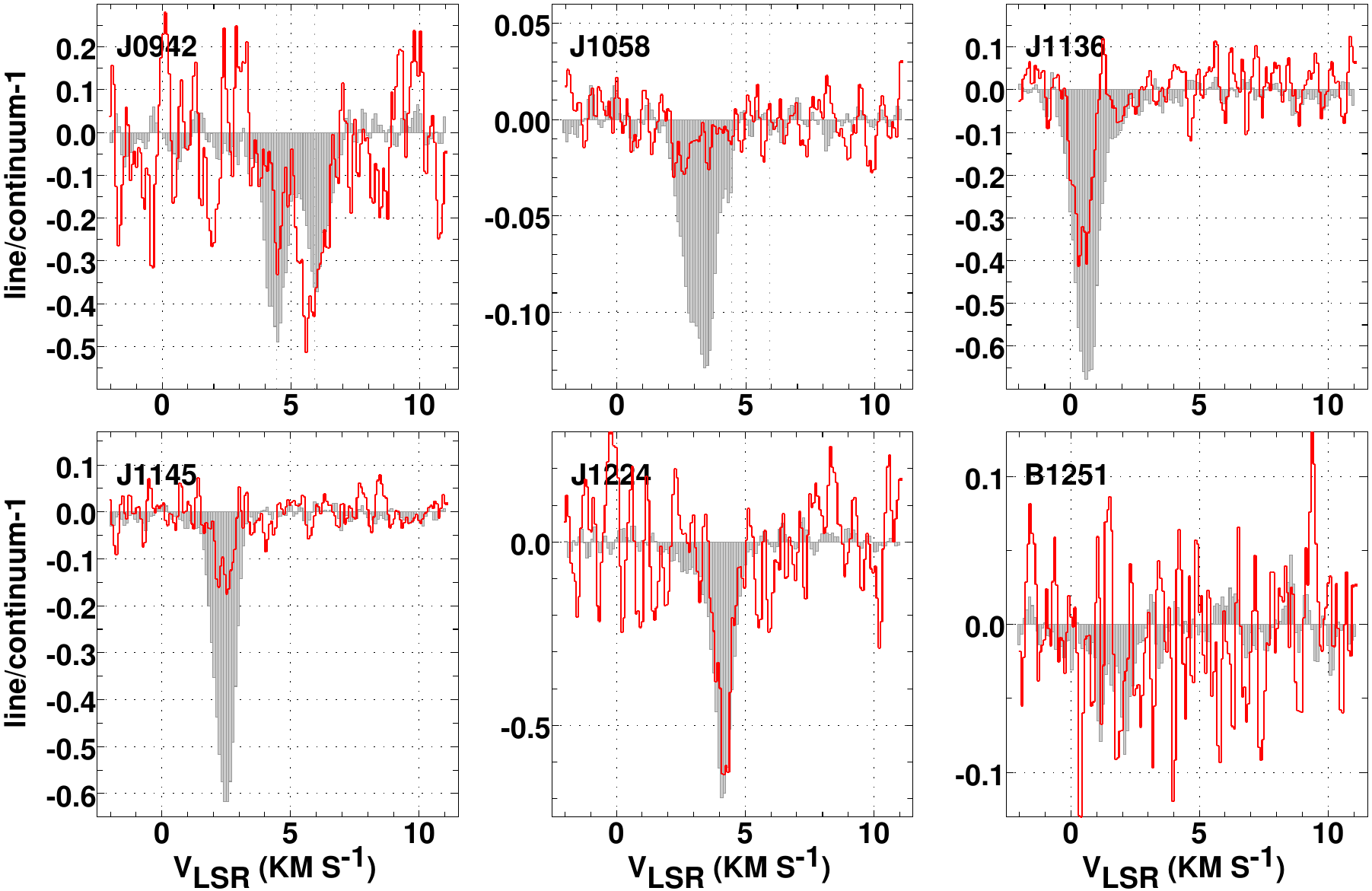}
  \caption[]{ALMA \hcop\ and \cotw\ absorption profiles for the six sources observed 
here.  The \hcop\ profile from Paper I is shown shaded in grey, the \cotw\ J=1-0 
profile from this work is shown in red.}
\end{figure*}

\subsection{New ALMA absorption measurements}

We observed the J=1-0 lines of \cotw\ and $^{12}$CN in absorption toward the six continuum 
sources whose line of sight properties are given in Table 1; the resulting CO line
profiles are shown in Figs. 1 and 2. The work was conducted 
under ALMA Cycle 5 project 2017.1.00227.S 
whose pipeline data products were delivered in 2018 February. The spectra discussed 
here were extracted from the pipeline-processed, continuum-subtracted data cubes at 
the pixel of peak continuum flux in the continuum map made from each spectral window, 
and divided by the continuum flux in the continuum map at that pixel. No (further)
baselining of any of the spectra was performed. Fluxes at 115.3 GHz ranged from 0.13 Jy 
for J1224 and 0.140 for J0942 to 1.68 Jy for J1058: several of the sources were 
considerably dimmer than in our Cycle 4 \hcop\ observations.  Each spectrum 
consisted of 1919 semi-independent channels spaced by 30.503 kHz corresponding to 0.079 
\kms\ at the 115.271 GHz rest frequency of \cotw.  The channel spacing is half the
spectral resolution.   CN was very tentatively detected in one direction, 
toward J1224, and will not be further discussed here.

\subsection{Other observational data at radio and UV wavelengths}

In Table 1 we quote the integrated \hcop\ optical depths from Paper I and the 
optically thin H I column densities derived from Galactic All-Sky Survey 
(GASS) III $\lambda$21 cm H I emission spectra 
\citep{KalHau15}. Column densities of \hcop\ are taken from Paper I as derived 
assuming rotational excitation in equilibrium with the cosmic microwave background, 
N(\hcop) $= 1.10 \times 10^{12} \pcc$ \Ihcop\  for a permanent dipole
moment $\mu =3.89$ Debye.   Fig. 1 shows the NANTEN CO J=1-0 emission spectra used
by \cite{Pla15Cham} to derive the column density of DNM along with the 
CO and \hcop\ absorption profiles taken during the course of our work that are 
shown in more detail in Fig. 2.

Fig. 3 shows the N(CO)-\ICO\ relation derived from the prior observations of J=1-0 
and J=2-1 CO emission and absorption by \cite{LisLuc98}: this is used to convert
our new measurements of \ICO\ to N(CO) as discussed in Sect. 4.1.  Fig. 4
compares CO and \HH\ column densities at radio and UV wavelengths using
the results of \cite{BurFra+07}, \cite{SonWel+07} and \cite{SheRog+08}.  

\subsection{Reddening and dust optical depth}

The 6\arcmin\ resolution dust-emission maps scaled to optical
reddening \EBV\ by \cite{SchFin+98} are used in Tables 1 and 4 . These 
reddening values can be converted to Planck 353 GHz dust optical 
depth \taup\ with errors of order $\pm6$\% using the relationship established by 
\cite{PlaXI} between \taup\ and reddening determined photometrically toward 
quasars, \EBV/$\tau_{353} = (1.49\pm0.03) \times 10^4$ mag.

\subsection{Conversion from integrated optical depth to column density for CO}

Here we measure the integral of the optical depth of the CO J=1-0 transition \ICO\ 
and additional assumptions are required to convert this to total CO column density.
Most generally, the integrated optical depth of the \cotw\ J,J+1 transitions are 
related to the column densities in rotational levels J and J+1 and the Einstein
B-coefficients as

$$\int \tJ d\nu = (h\nu/c)[N_JB_{J,J+1} - N_{J+1}B_{J+1,J}]  \eqno(1a) $$

For a linear rotor like CO with a permanent dipole moment $\mu = 0.11$ Debye 
(1 Debye = $10^{-18}$ esu) this can be recast as 

$$ N_J = {(2J+1)\over(J+1)} {{6.61\times10^{14}\pcc \tJ  
\over {(\mu/0.11D)^2 {(1-\exp(-h\nu/k\Tex))}}}}   \eqno(1b) $$

where $\int \tJ ~\rmdv$ is in units of \kms\ and  \Tex\ is the excitation 
temperature between levels $J$ and $J+1$, 
$N_{J+1}/N_J = (g_{J+1}/g_J)\exp(-h\nu/k\Tex)$ and $g_J = 2J+1$.

CO excitation temperatures are directly observed in diffuse molecular gas toward 
early-type stars \citep{BurFra+07,SonWel+07,SheRog+08} and for the J=1-0 transition
these are typically in the range 3-5 K (see Fig. 5 of \cite{Lis07CO} and Table 2 of 
\cite{Gol13}).  Similar 
values are inferred for gas observed in mm-wave CO absorption \citep{LisLuc98}.  In 
the  simple case of LTE at a single excitation temperature $\Tex$ for all J, the 
total column density can be written N(CO) = b \ICO, leading to 
b = $1.071 \times 10^{15} \pcc$ (\kms)$^{-1}$ for $\Tex = 2.73$K (the lower limit).
Other cases include b $= 1.381 \times 10^{15} \pcc$ (\kms)$^{-1}$
for $\Tex = 3.5$ K, b $= 1.614 \times 10^{15}\pcc$ (\kms)$^{-1}$ 
for $\Tex = 4$ K and b $= 1.871 \times 10^{15} \pcc$ (\kms)$^{-1}$ 
for $\Tex = 4.5$ K.

Here we convert the observed values of CO J=1-0 optical depths to column density using 
the earlier results of \cite{LisLuc98} who determined N(CO) along similar sightlines
using J=1-0 and J=2-1 observations of CO in absorption and emission.  Their results
for the integrated J=1-0 optical depths and CO column densities are very well represented 
with a slightly super-linear power-law regression fit as discussed in Sect. 4.1.  Given 
the numerical
examples in the preceding paragraph, this is roughly equivalent to assuming
an excitation temperature of 4.4-4.5 K for the stronger lines observed here (Table 1).   
An excitation temperature of at least 4.0 K is required to produce a 1 K brightness 
in CO J=1-0 emission although a smaller excitation temperature can produce an
integrated brightness \WCO\ $>$ 1 K-\kms\ with a sufficiently wide line. 

\begin{figure}
\includegraphics[height=8.1cm]{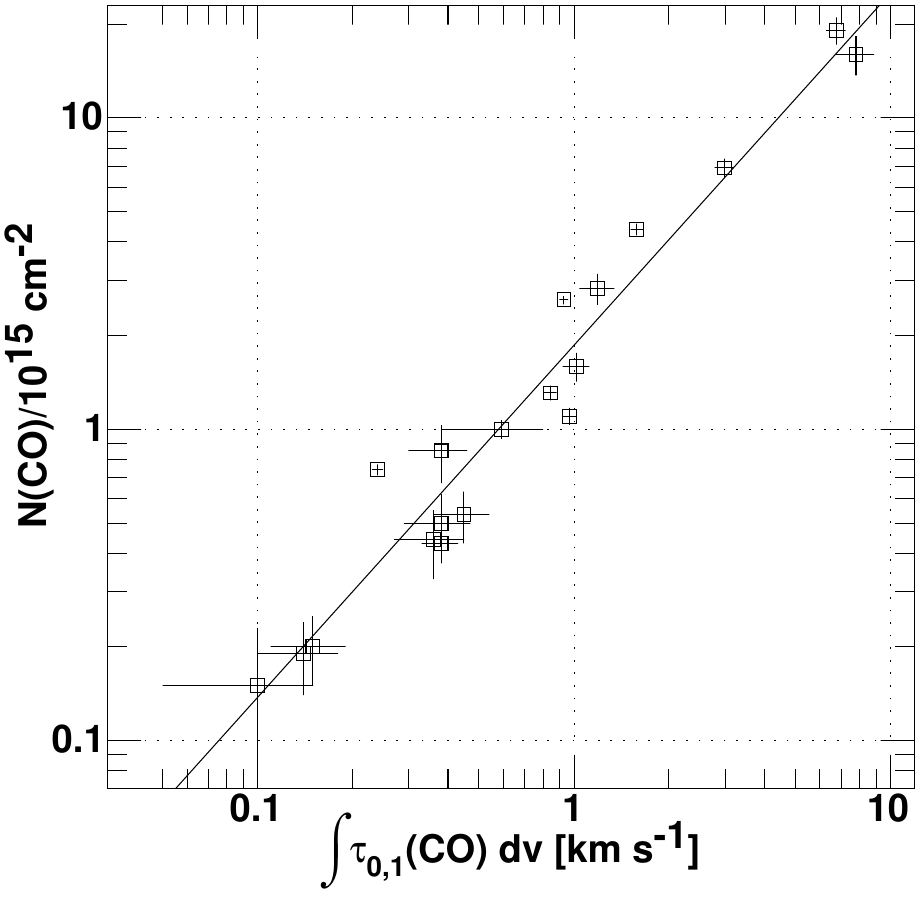}
\caption{The relationship between integrated CO J=1-0 optical
depth and N(CO) for the
data of \cite{LisLuc98} who observed J=1-0 and J=2-1 CO emission and
absorption along comparable sightlines.  The power-law fit is
N(CO) $=1.861\times 10^{15}\pcc \ICO^{1.131}$ where
\ICO\ $= \int \t01({\rm CO}~\rmdv$ in units of \kms .}
\end{figure}

\begin{figure}
\includegraphics[height=8.1cm]{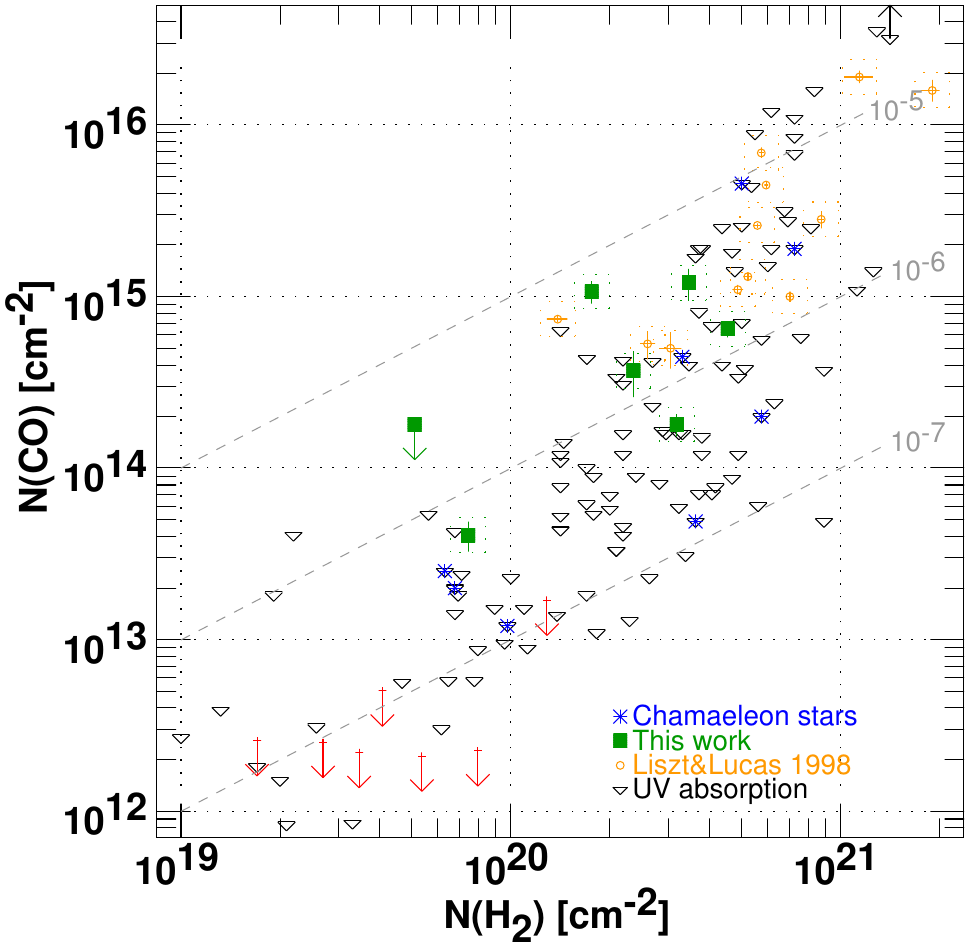}
\caption{\HH\ and CO column densities from UV and radio absorption spectra.
Shown are UV absorption line measurements from \cite{SonWel+07}, \cite{BurFra+07} 
and \cite{SheRog+08} (downward-pointing triangles),
mm-wave results from \cite{LisLuc98} (small orange circles),
and results from this work (green filled squares).  Some informative
upper limits to UV measurements are shown in red.  The mm-wave
results assume N(\HH) = N(\hcop)/$3\times10^{-9}$ and the present
results use Eqn. 2 to derive N(CO).  UV absorption
sightlines in Chamaeleon are marked by blue asterisks, see Table 4.
Grey dashed lines show the locii of relative abundances N(CO)/N(\HH) as
indicated. Only sighlines with N(CO) $ > 10^{15}\pcc$ are observable
in present wide-field CO emission surveys.}  

\end{figure}

\begin{figure}
\includegraphics[height=8.1cm]{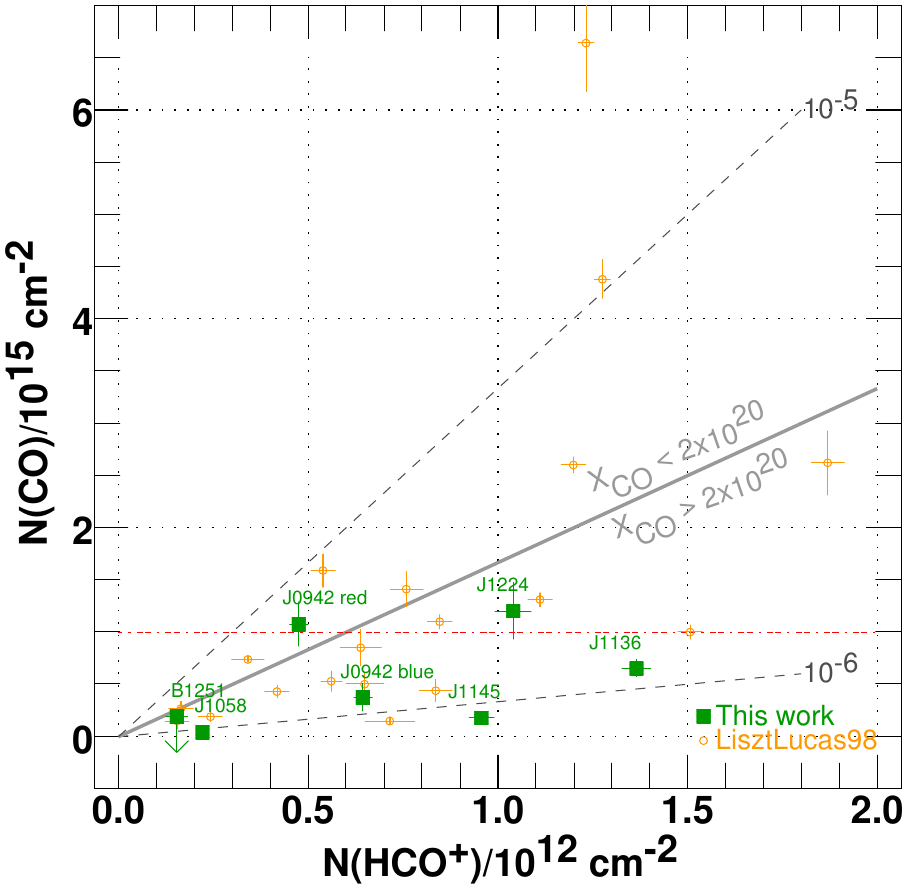}
\caption{CO and \hcop\ column densities.  Shown are results from this work 
(larger green filled squares) and the earlier observations of \cite{LisLuc98}
(small orange circles). Integrated CO J=1-0
optical depths \ICO\ have been converted to CO column density using
the relationship between $\Upsilon_{\rm CO}$ and N(CO) shown in Fig. 3,
N(CO) $=1.861\times 10^{15}\pcc \ICO^{1.131}$.  Grey dashed shaded
lines show fiducial values N(CO)/N(\HH) $ = 10^{-6}$ and $10^{-5}$ assuming
N(\hcop)/N(\HH) $=3\times 10^{-9}$. The broad grey line at N(CO)/N(\HH) 
$ = 5\times 10^{-6}$ divides the plot area into regions where \XCO\
is larger or smaller than $ 2 \times 10^{20}$ \HH\ (K-\kms)$^{-1}$.
The red dash-dotted line at N(CO) $= 10^{15}\pcc$ shows the CO column density
at which the integrated J=1-0 emission brightness \WCO\ = 1 K-\kms.}
\end{figure}

\begin{figure}
\includegraphics[height=8.1cm]{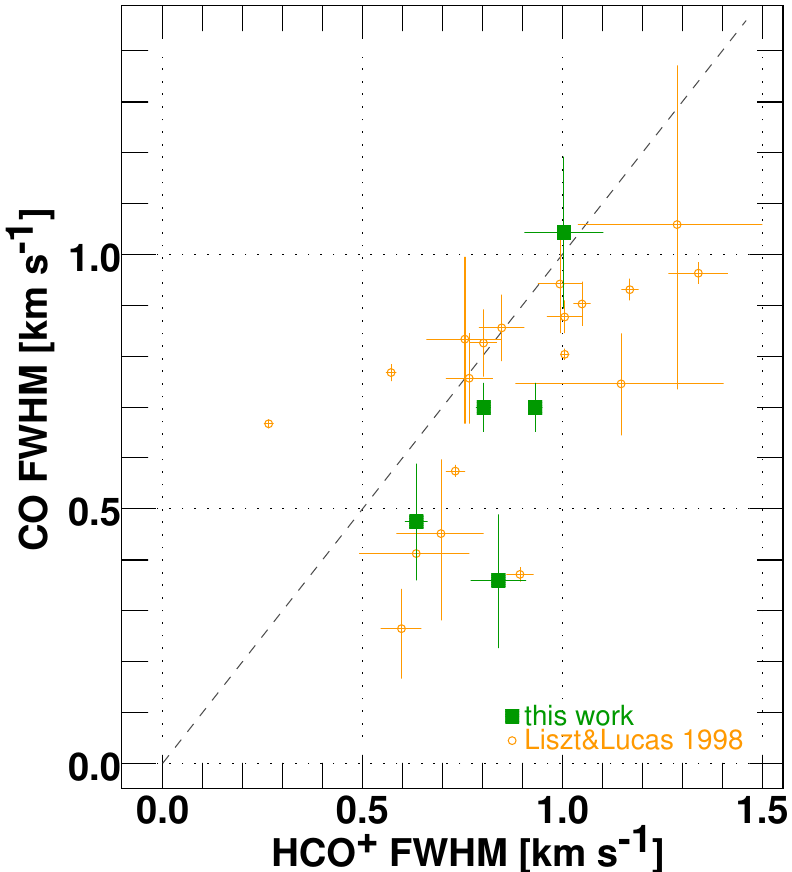}
\caption{CO and \hcop\ line profile FWHM. Shown are results from this work 
(larger green filled squares) and the earlier observations of \cite{LisLuc98}
(small orange circles).  The dashed gray line represents equality of the FWHM.}
\end{figure}

\begin{figure}
\includegraphics[height=8.1cm]{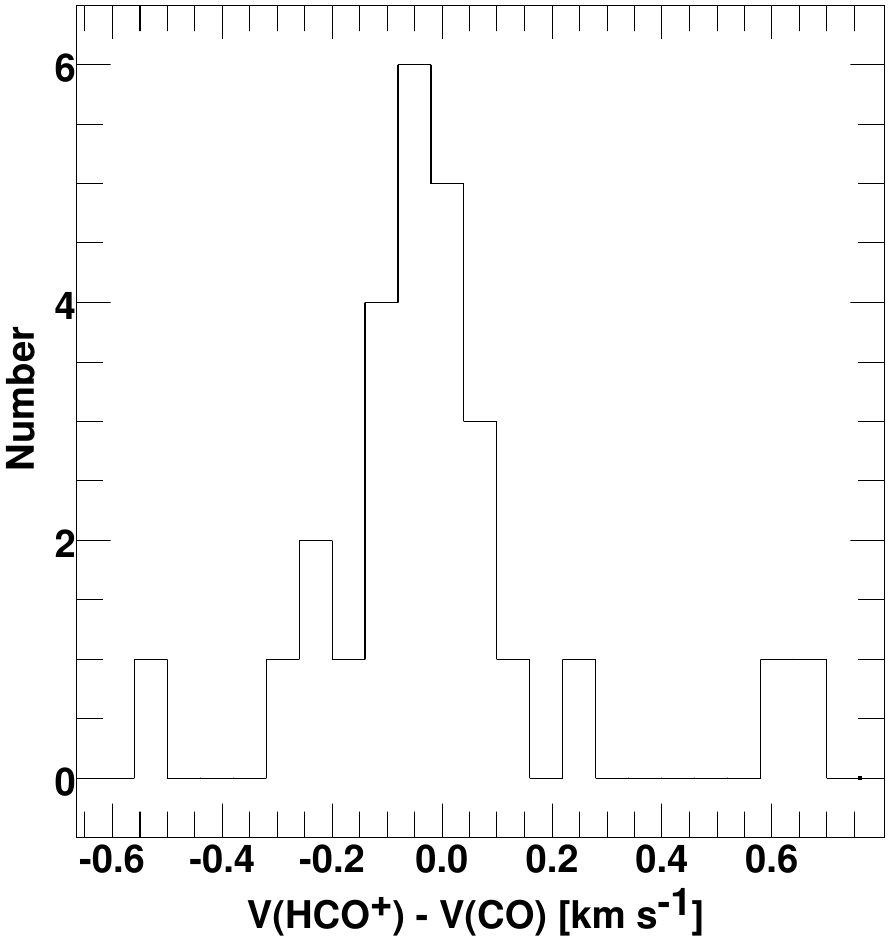}
\caption{Distribution of velocity differences between \hcop\ and CO kinematic
components from this work and the earlier observations of \cite{LisLuc98}.
The mean difference and its dispersion are 0.10 \kms\ $\pm$ 0.40 \kms .}
\end{figure}

As discussed by \cite{LisLuc98} and \cite{LisPet12}, the excitation temperature 
of the CO J=1-0 transition scales directly as the ambient thermal pressure of 
\HH\ and increases with  $\t01$. With a J=1-0 excitation temperature of 4.5 K
the implied partial pressures of \HH\ are n(\HH)\TK\ = $5.8\times10^3$ and 
$4.0\times10^3 \pccc$K for $\t01({\rm CO})$ = 0 and 1, respectively.  
These are typical values for the diffuse ISM sampled in C I by \cite{JenTri11}.

\subsection{The relative abundance of \hcop}

As in Paper I we use the relative abundance N(\hcop)/N(\HH) $ = 3\times 10^{-9}$
to derive N(\HH), the \hcop/\HH\ ratio being assumed to be constant in the gas.  
This value originates in the tight relationships that
exist between \hcop\ and OH \citep{LisLuc96,LisLuc00} or CH \citep{GerLis+19},
both of which are directly measured in optical/UV absorption spectra to have
fixed abundances relative to \HH\ \citep{Lis07CO,SonWel+07,SheRog+08,WesGal+09,
WesGal+10}. The overall uncertainty in the abundance of \hcop\ relative to \HH\ 
was estimated by \cite{GerLis+19} to be $\pm 0.2$ dex; see their
Table 4 giving relative abundances and uncertainties for a variety
of molecules used as proxies for \HH\ at radio/sub-mm wavelengths. 

\section{Observational results}

General properties of the sightlines observed here are given in Table 1,
including \hcop\ and CO optical depths integrated over the velocity range of 
the \hcop\ absorption.  Table 2 relates derived values of N(CO), N(\hcop) and 
N(\HH) as shown in Figs. 4 and 5.  Table 3 shows gaussian decompositions of 
the CO profiles in those cases where it is justified, compared with the profile
integral and results of gaussian decomposition of the \hcop\ profiles.  

Figure 1 is a finding chart for the thirteen sources observed in \hcop\ absorption, 
showing their placement on the sky with respect to the distributions of N(H I), 
N(DNM) and and the integrated CO brightness \WCO\  as derived by \cite{Pla15Cham}. 
Sightlines toward 6 of these were observed in CO, and CO and \hcop\ profiles in 
these directions are shown in panels arranged around the periphery of the chart
along with profiles toward J1723 that has detectable CO emission but was not observed 
in CO absorption in this work.

\hcop\ and CO absorption profiles are shown in Fig. 2 for all the sources observed 
here. CO absorption was detected along all five of sightlines in the high-DNM group
and not toward B1251.  The sightline to J0942 has the strongest integrated CO absorption 
in total but the strongest single feature is at 4 \kms\ toward J1224. The detection of 
very weak CO absorption toward J1058 was only possible because of its very strong 
continuum flux. 

The relative abundances of N(CO) and N(\HH) are known to have very wide variations 
resulting from their sensitivity to resonant photodissocation (Fig. 4) 
so it is not surprising that the ratio of the integrated CO and \hcop\ optical 
depths ranges from 0.7 to 7 among the sightlines measured here (Tables 1 and 3).  
A mild surprise is that 
CO absorption 
is so weak in the blue-shifted component having much stronger \hcop\ absorption 
toward J0942.  The NANTEN spectrum toward J0942 (Fig. 1) suggested the possible 
presence of CO emission overlapping the \hcop\ absorption and with stronger 
emission to the blue, in the same sense as that of the \hcop\ absorption.  This 
work shows that the CO column density in the blue-shifted component toward J0942 
is too small to support a CO detection in emission. 

\section{CO and \HH\ column densities, CO brightness and CO-\HH\ conversion
factor}

\subsection{\ICO, N(CO) and \WCO}

Figure 3 shows the data of \cite{LisLuc98} leading to the regression relation that 
we use to convert the CO J=1-0 integrated 
optical depth \ICO\ to column density N(CO) as discussed in Sect. 2.5.

$$ {\rm N(CO)} = 1.861\times 10^{15}\pcc \ICO^{1.131} \eqno(2) $$

Table 2 gives the results of using Eqn. 2 and N(\HH) = N(\hcop)/$3\times10^{-9}$
to derive N(CO) and N(\HH) from our mm-wave CO and \hcop\ absorption measurements.
As a check on this procedure, Fig. 4 compares the present and prior mm-wave results 
for N(CO) and N(\HH) with the larger sample of direct UV absorption line measurements
of \cite{SonWel+07}, \cite{BurFra+07} and \cite{SheRog+08}. Results for UV sightlines 
in the Chamaeleon region are noted in Fig. 4 (see Table 4) and are not exceptional. 
  
The mm-wave results appear typical, not outliers, lending confidence to the 
assumed relative abundance of \hcop\ that was used to derive N(\HH).  The mean 
relative abundance $<$N(CO)/N(\HH)$> \simeq 2.2-2.5\times 10^{-6}$ for the lines of 
sight observed here (Table 2) is very small compared to the expected fractional 
abundance of free gas-phase carbon, N(C)/N(H) $\simeq 1.6 \times 10^{-4}$ 
\citep{SofLau+04}, but still larger than found along many of the 
sightlines represented by UV absorption in Fig. 4.  The older mm-wave results of 
\cite{LisLuc98} have CO and \HH\ column densities toward 
the upper ends of their respective distributions in Fig. 4, and stronger 
\coth\ fractionation than is typically the case for the UV measurements \citep{Lis17Frac}. 

In Fig. 5 we compare N(CO) and N(\hcop) from the present work with those in the 
earlier study of CO in emission and absorption \citep{LisLuc98}.  
The smaller sample of new data lacks individual features with column densities 
N(CO) $> 2\times 10^{15}\pcc$ or  N(\hcop) $> 1.5 \times 10^{12}\pcc$ as found in the 
older data  but results for the two datasets largely overlap.

\subsection{The implied CO-\HH\ conversion factor}

Also plotted in Fig. 5 are dashed lines  corresponding to relative abundances 
N(CO)/N(\HH) =  $10^{-6}$ and $10^{-5}$, which follow directly from 
the relative abundance N(\hcop)/N(\HH) =  $3\times 10^{-9}$. 
An implied CO-\HH\ conversion factor can be derived when it is recognized that
the integrated J=1-0 emission profile brightness \WCO\ is related to
N(CO) as \WCO/(1 K-\kms) = N(CO)/$10^{15} \pcc$ over a wide range
of hydrogen number density in diffuse molecular gas \citep{Lis07CO,Lis17Frac}, 
maintaining a high degree of linearity up to brightnesses \WCO\ $\la 5-10$ K-\kms .  
In this case the CO-\HH\ conversion factor depends only on the relative
abundance N(CO)/N(\HH) and the relative abundance 
N(CO)/N(\HH) $= 5\times 10^{-6}$ divides Fig. 5 (and the rest of the very local 
Universe) into regions where the CO-\HH\ conversion factor is larger or smaller
than $2\times 10^{20}\pcc$ (K-\kms)$^{-1}$.  That this value of the CO-\HH\
conversion factor is reached at such low CO/\HH\ relative abundances is a
consequence of the very high brightnesses per molecule that occur  when the
rotational excitation is so strongly sub-thermal \citep{GolKwa74,LisPet+10}.

Only in the red-shifted component toward J0942 is the implied CO-\HH\ conversion factor 
N(\HH)/\WCO\ below $2\times 10^{20}~\HH \pcc$ (K-\kms)$^{-1}$.
We did not observe CO absorption toward J1723, but consideration of the detected
CO emission and \hcop\ absorption in this direction (Fig. 2) provides an interesting 
point of comparison.  Directly integrating over the separate velocity ranges of the 
two kinematic components results in CO integrated brightnesses of \WCO = 0.90$\pm$0.14
and 1.36$\pm$0.14 K-\kms\ and \hcop\ integrated optical depths \Ihcop = 0.70$\pm$0.03
and 0.81$\pm$0.03 \kms.  Converting \Ihcop\ to N(\hcop) and
N(\HH) = N(\hcop)$/3\times10^{-9}$ leads to CO-\HH\ conversion factors  
N(\HH)/\WCO\ = 2.8$\pm0.5$ and 2.2$\pm 0.2 \times 10^{20}~ \HH \pcc$ (K-\kms)$^{-1}$ 
or N(\HH)/\WCO\ $= 2.4\pm0.3 \times 10^{20}~ \HH \pcc$ (K-\kms)$^{-1}$ overall.

\subsection{Kinematics of CO absorption features}

In Figs. 4 and 5 we showed that the column densities of CO and \hcop\ in the outskirts 
of Chamaeleon are like those previously observed along random lines of sight through
the diffuse ISM generally and the same is true of the linewidths.
In Table 3 we show the results of gaussian decomposition of the \hcop\ and CO line
profiles that are shown in Fig. 2.  In some cases the superior signal/noise of the 
\hcop\ measurements allows fitting of wings and closely-separated components that 
is not feasible in CO.  The plotted results represent only those features that
can be identified and associated in both species and have been
corrected  by quadrature differencing of the FWHM and velocity resolution.
The FWHM linewidths for CO are systematically smaller than those of \hcop\ as 
shown in Fig. 6. The unweighted mean linewidths for the combined
sample of new and old data are $0.87\pm0.24$ \kms\ for 
\hcop\ and $0.72\pm0.22$ \kms\ for CO. This is not necessarily expected since 
CO forms from the recombination of \hcop\ in diffuse molecular gas 
\citep{GlaLan76,Lis07CO,VisVan+09} and is initially cospatial with parent 
\hcop\ molecules.  \hcop\ and OH profiles do not show comparable differences 
\citep{LisLuc00}. However, recombination
of \hcop\ to CO is sensitive to temperature while survival of 
the CO molecules after formation depends on shielding by \HH, dust and other 
CO molecules \citep{DraBer96,VisVan+09,sternberg:14,Lis15HD},raising the 
possibility that regions with differing N(\hcop)/N(CO) exist within the gas 
and are superposed along the line of sight \citep{GodFal+09}. 

Linewidth differences among other chemical species are well-known in diffuse 
molecular gas. In optical spectra there is a sequence CH\p, CH, CN from 
broader to narrower lines \citep{LamCra+90,CraLam+95,PanFed+05} perhaps 
extending further to CO \cite{PanFed+05}. At mm-wavelengths, CN absorption
lines are narrower than those of HCN and HNC \citep{LisLuc01,GodFal+10} from 
which CN forms by photodissociation, and all the CN-bearing species have
narrower lines than in \hcop.  CH\p\ lines are typically much wider than 
those of other molecules, presumably because the formation of CH\p\ occurs
through the strongly endothermic reaction of C\p\ and \HH.  \hcop\ may keep a 
memory of this excess kinetic energy seen in the CH\p\ linewidth because it 
is such a close daughter product of CH\p\ \citep{GodFal+14}. 

Linewidth differences could arise as the result of the interplay between
chemistry and gas dynamics in the diffuse molecular gas but 
not because CO molecules in the line wings of mm-wave absorption profiles
are more susceptible to 
photodissociation.  The shielding and self-shielding of \HH\ and CO arise from 
the continuous opacity of dust and the wings of strongly-damped \HH\ and CO 
electronic transitions at the column densities needed to form the observed amounts 
of CO, and the Doppler velocity distribution of CO or \HH\ molecules is not 
important to their survival.

In Paper I we noted a kinematic segregration in the Chamaeleon gas in the sense 
that the directions with less DNM and small N(\HH) were seen in \hcop\ only at 
v $\la 2$\kms\ while the material with velocities 2 - 6 \kms\ had the preponderance 
of the HCN absorption, stronger H I emission and much more DNM and \HH.  We suggested 
that this might indicate the presence of slow shocks in the region. The CO data in
Chamaeleon are too limited to draw firm conclusions but Fig. 7 shows
a histogram of the distribution of line centroid velocity differences between
\hcop\ and CO for the new and old data.  The mean difference is 0.1 \kms, with a
dispersion of .4 \kms, indicating at most a very small systematic shift, and a 
small but physically significant dispersion. 

Dynamical processes are important for the formation chemistry of the molecules 
we observe, most obviously for \hcop\ whose observed abundance in diffuse molecular 
gas N(\hcop)/N(\HH) $ = 3\times 10^{-9}$ is nearly 100 times higher than would 
be expected from Langevin rates for the reaction of C\p\ + OH with their observed 
column densities \citep{VanBla86}. \cite{FloPin99} and \cite{LesPin+13}
discussd how a collection of 
slow shocks in magnetized diffuse gas could produce CH\p, \hcop\ and CO. Most models 
that boost the CH\p\ and/or \hcop\ chemistry in diffuse gas tap the energy 
of an ion-neutral drift \cite{ValGod+17} which has a broad distribution ranging from 
0.01 to 1 km/s.  \cite{VisVan+09} drew on an earlier model of \cite{FedRaw+96}, calculating 
molecular abundances at a suprathermal effective temperature determined by an Alven 
velocity of 3.5 \kms.

Forms in localized regions (vortices) where turbulent energy is dissipated and the ion-neutral drift is maximum.

The turbulent dissipation region (TDR) model of diffuse cloud chemistry 
\citep{GodFal+14} draws on earlier work \citep{FalPin+95,JouFal+98,PetFal00}
and forms CH\p\ and \hcop\ in highly localized regions (vortices) where turbulent 
energy is dissipated and the ion-neutral drift is maximum.  \hcop\ subsequently 
relaxes and accumulates at the observed abundance 
in ambient gas where it thermalizes and recombines to CO. CO is subsequently 
fractionated by C\p\ exchange at ambient kinetic temperature \citep{Lis17Frac}
and both \hcop\ and CO are observed with a degree of rotational excitation that 
is characteristic of the typical thermal pressures for the diffuse interstellar 
medium at large \citep{JenTri11}, as noted in Sect. 2.5.  In such a model
\hcop\ recombines to CO more rapidly as it thermalizes, and the \hcop-CO linewidth
difference is presumably a collateral effect. 

Comparable linewidth differences are expected between \hcop\ and other species that 
form by the recombination of ions, for instance HCN, HNC and CN that form from recombination 
of the as-yet undetected species HCNH\p.  \cite{GodFal+10} observed smaller linewidths
in HCN and HNC compared to \hcop\ (\cite{LisLuc01} did not), and there is general 
agreement that the smallest linewidths have generally been observed in CN.  We sought 
but did not detect CN except marginally toward J1224.  The difference in the velocity 
profiles therefore appear to be related to the interplay of chemical and dynamical 
processes controlling the formation of molecules. To explore further the origin of the 
difference in velocity profiles more data are needed in specific regions where the 
large-scale orientation of the gas flows and magnetic field are homogeneous.

\section{Molecular, missing, and CO-dark gas}


 In Paper I we showed that \hcop\ absorption lines traced important \HH\ reservoirs 
extending several parsecs beyond the CO-bright parts of the Chamaeleon molecular clouds. 
The \HH\ column densities inferred from the \hcop\ data for a standard abundance
ratio N(\hcop)/N(\HH) $= 3\times10^{-9}$ could substantially account for 
the excess gas along the eight sight lines with large DNM column densities 
N(DNM) $\ga 2\times10^{20}\pcc$: diffuse \HH\ partially contributed to the DNM toward 
three directions and fully explained the DNM toward the five densest directions. 
In this section we extend some of the analysis from Paper I using an 
expanded, hybrid sample based on direct UV absorption measurements of N(\HH) and 
N(CO).  

In Figs. 1 and 4 we noted the existence of UV absorption measurements 
of N(\HH) and N(CO) toward stars of widely-varying distance seen toward 
Chamaeleon (Table 4). The nearer of these stars could in principle lie in 
front of some portion of the Chamaeleon gas but N(\HH) is actually
relatively high for the nearer objects.  To complement the modest 
sample of sightlines 
observed with ALMA, we extracted the HI and DNM column densities and mm-wave
CO brightness W(CO) toward these stars and combined this information with 
that for the mm-wave sightlines to produce Fig. 8 (like Fig. 6 in Paper I). 
In Fig 8 we compare the total hydrogen 
column density obtained from the dust and gamma-ray data (the x-axis) to
the total hydrogen column density N(HI)+2N(\HH) derived with and without the 
use of CO emission. In the left panel N(\HH) = \XCOa \WCO, while at right
N(\HH) is measured indirectly from \hcop\ absorption lines for our ALMA data 
(circles) or directly from UV \HH\ absorption lines (crosses). The error bars 
on N(\HH) at right assume $\pm25$\% errors in N(\HH) from the UV data and 
$\pm58$\% errors on the N(\hcop)/N(\HH) ratio \citep{LisGer16}). 

The colour coding with
DNM column density shows that diffuse \HH\ can fully account for the DNM column 
densities over a range varying by at least a factor of six.  One can also note from 
the difference in column density scales between the total and DNM gas that the broad 
atomic envelopes of the clouds contribute more gas along those sight lines than the 
diffuse \HH\ reservoirs. As in Paper 1 we infer that the DNM is  mostly molecular
while the gas overall is atomic, for the lines of sight we studied outside the 
CO-bright portions of the Chamaeleon complex.

\begin{figure*}
\includegraphics[height=9.7cm]{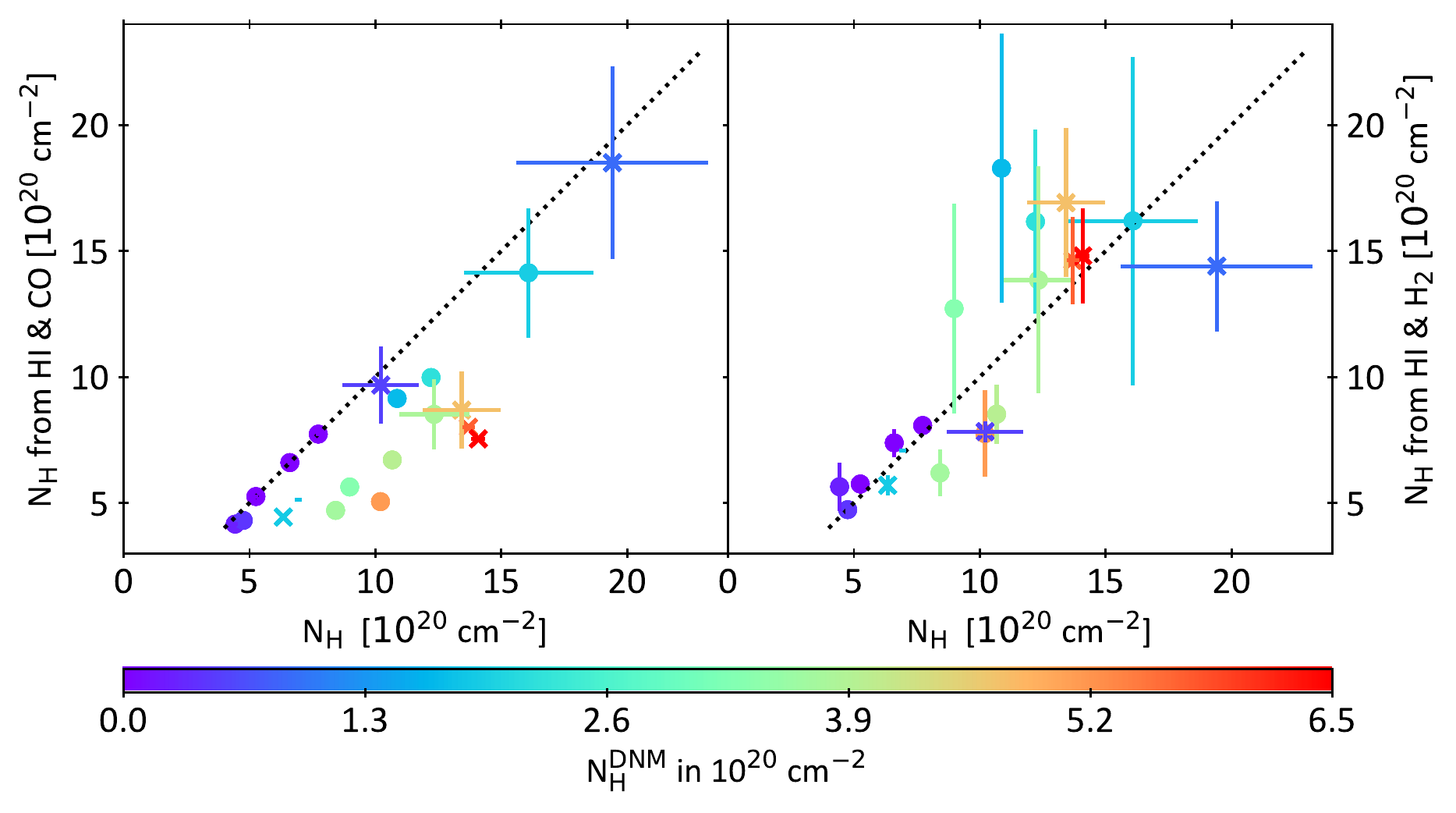}
  \caption[]{Comparison of total column density of H-nuclei N(H) derived from 
the DNM analysis and from observable column densities.  The variable plotted 
along the horizontal axis, N(H) = N(H I)$|_{\rm cham}$+N(DNM)+2\WCO \XCOa, is 
the quantity fitted in the global DNM analysis \citep{Pla15Cham} and is the same 
in both panels. The vertical axis in each panel is the inferred total hydrogen 
column density N(H) = N(H I)$|_{\rm cham}$+2N(\HH) but with N(\HH) taken differently 
at left and right.  At left, N(H) is calculated with N(\HH) = \WCO \XCOa. At right, 
N(H) is calculated with N(\HH) = N(\hcop)$/3\times10^{-9}$ for the sightlines
observed in \hcop\ absorption, or using actual measured N(\HH) for the stars 
with measured N(\HH) and N(CO) (see Fig. 4 and Table 4).  The error bars assume 
$\pm25$\% errors in N(\HH) from UV absorption and $\pm58$\% errors in the 
N(\hcop)/N(\HH) ratio. Stars are marked by crosses and mm-wave targets by filled
circles, and all sightlines are color-coded according to N(DNM) as shown at
bottom.}
\end{figure*}

\section{Summary}

MM-wave CO absorption line measurements are well-suited to detecting or
constraining the properties of  dark medium and dark CO.
The CO absorption measurements performed here answer the question of why the 
CO emission fails to represent N(\HH) reliably in the outskirts of Chamaeleon.
CO column densities and CO relative abundances are small in diffuse molecular gas 
around the HI $\rightarrow$ \HH\ transition as the result of the CO chemistry at 
moderate reddening.

In Paper I we detected  mm-wave J=1-0 absorption from \hcop\ and, less commonly, 
\cch\ and HCN, against thirteen compact extragalactic continuum sources seen 
toward the outskirts of the Chamaeleon HII region-molecular cloud complex.  
Only one of these directions (J1723) had a firm detection of CO J=1-0 emission.  
Eight of the thirteen directions comprised a group with higher column densities 
of dark neutral medium and, as we showed, much higher molecular column densities.
Converting the N(\hcop) to N(\HH) with N(\hcop)/\N(\HH) $= 3\times 10^{-9}$ 
we found that the DNM was mostly molecular even while the gas as a whole (in the
observed directions) was predominantly atomic, with H-nuclei in atomic form.

In this work we observed CO J=1-0 absorption lines toward six of the thirteen 
sightlines from Paper I, five of them in the higher-DNM group (Table 1 and Fig. 1).  
All five of these directions were detected in CO absorption (Fig. 2). Using
a conversion between integrated J=1-0 optical depth and CO column density
derived earlier from observations of J=1-0 and J=2-1 absorption and emission
along comparable sightlines (see Eqn 2 and Fig. 3, and Sect. 4) the observed 
CO column densities were found to lie in the range 
$4\times 10^{13} \pcc \la $ N(CO) $\la 1.2\times 10^{15}\pcc$.
This is generally below the expected limit of detectability of extant CO emission 
surveys given that an integrated CO brightness of \WCO\ = 1 K-\kms\ corresponds to 
N(CO) $= 10^{15}\pcc$ in diffuse molecular gas (Sect. 4.2). As noted in Paper 1, 
the Planck CO emission foreground maps are, lamentably, hashed in the directions 
we observe, presumably as the result of point-source removal.

The ratio of J=1-0 integrated optical depths in CO and \hcop\ varies by almost
an order of magnitude (Table 1 and 2).  Comparably-wide relative abundance
variations are typical of sightlines at moderate reddening seen in UV absorption 
(Fig. 4) owing to the sensitivity of \HH\ and CO column densities to self- and 
mutual shielding and shielding by dust.  To check on the conversions we employed to 
derive N(\HH) and N(CO) we compared the radio- and UV-determined CO and \HH\ 
column densities in Fig. 4, finding the two types of measurements to yield
consistent determinations of the run of CO and \HH\ column density.

In Fig. 5 we compared N(CO) and N(\hcop) for the new and existing mm-wave CO 
absorption measurements, leaving the conversion to N(\HH) implicit.  The 
datasets are consistent and there is nothing unusual about the Chamaeleon 
measurements, although they lack sightlines with \hcop\ and CO column densities 
as high as some of those seen earlier (this is also apparent in Fig. 4). 
Given the equivalence of 
CO brightness and column density \WCO\ $\simeq$ N(CO)$/10^{15}\pcc$ in the 
diffuse molecular gas, the CO-\HH\ conversion factor N(\HH)/\WCO\ is 
linearly proportional to the CO relative abundance, and 
N(\HH)/\WCO\ $\le 2 \times 10^{20}~ \HH \pcc$ (K-\kms)$^{-1}$
only when N(CO)/N(\HH) $\ge 5\times 10^{-6}$.  This is more than
twice the mean value observed here: only the red-shifted line toward J0942 
has a CO relative abundance high enough that the implied CO-\HH\ conversion 
factor is below N(\HH)/\WCO\ $= 2 \times 10^{20}~ \HH \pcc$ (K-\kms)$^{-1}$
(see Table 2 and Fig. 5). High values of the CO-\HH\ conversion factor 
N(\HH)/\WCO\ were also inferred in Paper 1 based on the limits to the CO 
brightness.

The mm-wave CO absorption line chemistry has not been revisited in the
20 years since the observations of \cite{LisLuc98}, which is somewhat 
surprising given the kind of information that is returned.  For this 
reason, and owing to the modest number of sightlines we studied, we 
somewhat extended the discussion beyond the Chamaeleon region 
to a more general comparison of CO with \hcop\ and \HH. Beyond the column 
densities and abundances shown in Fig. 4 and 5, we noted (Fig. 6, Table 3) 
that the CO FWHM linewidths are consistently about 20\% smaller than those 
seen in the progenitor \hcop\ molecule ($0.72\pm0.22$ \kms\ and $0.87\pm0.24$ 
\kms\ respectively). A comparable difference is seen in CN-bearing species 
when compared with \hcop. There is also a $\pm0.4$ 
\kms\ dispersion in the difference of the centroid velocities of \hcop\ and CO 
(Fig. 7).  In Sect. 4 we discussed how the abundances and kinematics of 
\hcop, CO and other molecules arise from gas dynamical processes that enhance 
the formation rate of progenitor ions like CH\p\ and \hcop\ and
create inhomogeneities that cause linewidth differences.

In Sect. 8 we revisited part of the DNM analysis of Paper I, using a 
hybrid sample whereby we augmented the ALMA observations with the 
directly-determined \HH\ and CO column densities measured in UV absorption
toward 8 bright stars seen toward the Chamaeleon region.  As in Paper I 
we concluded that the DNM in the outskirts of the Chamaeleon complex is
mostly molecular, with most H-nuclei in \HH, while the gas overall is 
primarily atomic.

CO absorption profiles provide
more information than just the column density, and consideration of a wider range 
of CO absorption line observations should help to elucidate the formation of
molecules and the presence of DNM and the  ``dark'' gas whose shadows we explored
in this work.

\begin{table*}
\caption[]{Sightline and velocity-integrated spectral line properties}
{
\small
\begin{tabular}{lcccccccccc}
\hline
Source&$\alpha$(J2000)&$\delta$(J2000)&$l$&$b$&\EBV$^a$&N(H I)$^b$&\Ihcop$^c$&$S_{115}$&${\sigma_{l/c}}^d$&\ICO$^c$ \\
 & hh.mmssss & dd.mmssss &\degr & \degr & mag & $10^{20}\pcc$ & \kms & Jy &  & \kms \\ 
\hline

J0942-7731&09.424275 &-77.311158 &293.321 &-18.329 &0.33&9.1&1.142(0.067)$^e$ &0.140 &0.126 &0.855(0.133)\\
J1058-8003&10.584331 &-80.035416 &298.010 &-18.288 &0.15&6.0&0.201(0.009)  &1.684 &0.010& 0.034(0.008) \\
J1136-6827&11.360210 &-68.270609 &296.070 &-6.590  &0.47&21.7&1.241(0.035) &0.344&0.054& 0.396(0.047) \\
J1145-6954&11.455362 &-69.540179 &297.316 &-7.747  &0.38&16.8&0.870(0.031) &0.517&0.034& 0.125(0.025)  \\
J1224-8313&12.245438 &-83.131010 &302.095 &-20.391 &0.26&8.7 &0.945(0.044)        &0.132 &0.122 & 0.679(0.137) \\
B1251-7138&12.545983 &-71.381840 &303.213 &-8.769  &0.28&17.0&0.139(0.027)&0.323 &0.055 & $<$0.131$^f$ \\
\hline
\end{tabular}}
\\
$^a$ \cite{SchFin+98} \\
$^b$ N(H I) $= \int {\rm T}_{\rm B}~{\rm dv} \times 1.823 \times 10^{18}\pcc $ from the Gass III H I profile \citep{KalHau15} \\
$^c$ \Ihcop\ and \ICO\ are the integrated J=1-0 optical depth in units of \kms \\
$^d$ line/continuum rms at zero optical depth in the 2x oversampled spectrum \\
$^e$ Quantities in parenthesis are the standard deviation \\
$^f$ Upper limit is 3$\sigma$ \\

\end{table*}

\begin{table*}
\caption[]{\hcop\ \HH\ and CO column densities}
{
\small
\begin{tabular}{lcccccccc}
\hline
Source  & v$_0$ & N(\hcop)$^a$ & N(\HH)$^b$ & N(CO)$^c$         & N(CO)/N(\HH) \\
        & \kms\ & $10^{12} \pcc$ & $10^{20} \pcc$ & $10^{15}\pcc$ & $10^{-6}$    \\
\hline
J0942 & 4.5 & 0.64(0.02)$^d$ & 2.35 & 0.37(0.11) & 1.56 \\
      & 6.0 & 0.48(0.02) & 1.76 & 1.07(0.16) & 6.08 \\
\hline
J1058 & 3.2 & 0.22(0.04) & 0.74 & 0.04(0.01) & 0.55 \\
J1136 &  0.6 & 1.48(0.04) & 4.55 & 0.65(0.07) & 1.42 \\
J1145&2.5 & 0.96(0.03) & 3.19 & 0.18(0.03) & 0.56 \\
J1224 &4.0 & 1.04(0.05) & 3.47 & 1.20(0.26) & 3.45 \\ 
B1251 & 1.7 & 0.15(0.03) & 0.51 & $<0.18^e$ & $<3.66^e$ \\
\hline
mean$^f$ &3.4(1.8) & 0.80(0.45) & 2.68(1.35)& 0.59(0.47) & 2.27(2.15) \\
mean$^g$ & 3.2(1.8) & 0.71(0.48) & 2.37(1.48)& 0.53(0.46) & 2.47(2.02) \\
\hline
\end{tabular}}
\\
$^a$N(\hcop)$=1.10\times10^{12}\pcc  {\rm I}_\hcop$ \\
$^b$N(\HH) = N(\hcop)$/3\times10^{-9}$ \\
$^c$N(CO)$=1.861\times10^{15}\pcc {\rm I}_{\rm CO}^{1.131}$ \\
$^d$ quantities in parentheses are the standard deviation \\
$^e$ upper limits are $3\sigma$ \\
$^f$ for the six components with CO detections \\
$^g$ including B1251 at the upper limit \\
\end{table*}

\begin{table*}
\caption[]{Gaussian decomposition of \hcop\ and CO absorption}
{
\small
\begin{tabular}{lcccccccc}
\hline
 & \hcop & \hcop & \hcop &\hcop & CO & CO & CO &CO  \\
\hline
Source  & v$_0$  & $\tau_0$ & FWHM & \Ihcop  & v$_0$  & $\tau_0$ & FWHM  & \ICO  \\
        & \kms   &          & \kms & \kms             & \kms   &          & \kms  & \kms  \\
\hline
J0942  &4.478  &0.634 &0.846&0.571   &  4.494 & 0.347 & 0.368 & 0.136  \\
   $\pm$ &0.029 &0.046 &0.069 &0.040 &  0.056  & 0.118 & 0.128 & 0.043  \\
  &6.045  &0.402 &1.009 &0.432   &  5.766 & 0.568 & 1.046 & 0.632  \\
   $\pm$ &0.040 &0.035 &0.098 &0.036  &  0.064  & 0.086 & 0.149 & 0.085 \\
\hline
J1058   &3.236  &0.134&1.409 &0.201 &&&&  \\
   $\pm$ &0.014 &0.003 &0.035 &0.004 &&&&   \\
\hline
J1136 &  0.607 &  1.100 &  0.938 &  1.098  & 0.449 & 0.507 & 0.704 &  0.380 \\
   $\pm$ &  0.007 & 0.022  & 0.017  & 0.019    & 0.022 & 0.038 & 0.048 & 0.025  \\
         &  1.880  &  0.104  &  0.684 & 0.076     & & & &  \\
   $\pm$ &  0.042  & 0.011  & 0.101  &0.009   & &  &  \\
\hline
J1145&2.501  &0.938 &0.809 &0.808 &  2.426 &   0.171 &  0.701 &  0.128  \\
   $\pm$ &0.009 &0.030 &0.019 &0.021& 0.042  & 0.022  & 0.097  & 0.016  \\
\hline
J1224    &3.741 &0.127 &2.261 &0.307 & &&& \\
   $\pm$ &0.149 &0.025 &0.287 &0.047 & &&& \\
  &4.218  &0.979 &0.643&0.670  &  4.215 &  0.975 & 0.481  & 0.500  \\
   $\pm$ &0.012 &0.068 &0.028 &0.036 &  0.057  &  0.169 &  0.113  & 0.094  \\
\hline
\end{tabular}}
\\
\end{table*}

\begin{table*}
\caption[]{Stars toward Chamaeleon studied in UV absorption$^a$}
{
\small
\begin{tabular}{lccccccc}
\hline
Star  & l & b & distance & \EBV$_*$ & \EBV$_{\rm SFD}$$^b$ & N(\HH) & N(CO) \\
 HD/CPD   & \degr   & \degr & pc  & mag &  mag & $10^{20} \pcc$ &  $10^{14}\pcc$ \\
\hline
-69 1743 & 303.71 & -7.35 & 4700 & 0.30 &0.29 & 1.00 & 0.12 \\
93237& 297.18 & -18.39 & 310 & 0.09 & 0.11 & 0.63& 0.25  \\
94454&   295.60 & -14.73 & 330  & 0.18 & 0.31 &  5.75&2.00  \\
96675&   296.62& -14.57 & 160 & 0.30 & 0.64 & 7.24& 19.1  \\
99872&   296.69& -10.62 &230  &0.36 & 0.38 & 3.31&4.47  \\
102065$^c$&  300.03& -18.00 & 170 & 0.17 & 0.38 & 3.63&0.49 \\
116852$^c$&  304.08& -16.13 & 4800 & 0.21& 0.22 & 0.68 &0.20  \\
203532& 309.46 & -31.74 &250  & 0.28 & 0.34  & 5.00 & 45.71 \\
\hline
\end{tabular}}
\\
$^a$ Stellar quantities from \cite{SheRog+08} \\
$^b$ Reddening from \cite{SchFin+98} \\
$^c$ Also observed by \cite{BurFra+07} \\
\end{table*}

\begin{acknowledgements}

This paper makes use of the following new ALMA data: ADS/JAO.ALMA\#2017.1.00227.S .
ALMA is a partnership of ESO (representing its member states), NSF (USA)
and NINS (Japan), together with NRC (Canada), NSC and ASIAA (Taiwan), and
KASI (Republic of Korea), in cooperation with the Republic of Chile.  The
Joint ALMA Observatory is operated by ESO, AUI/NRAO and NAOJ.
The National Radio Astronomy Observatory is a facility of the National Science 
Foundation operated under cooperative agreement by Associated Universities, Inc. 

This work was supported by the French program “Physique et Chimie du Milieu 
Interstellaire” (PCMI) funded by the Conseil National de la Recherche Scientifique 
(CNRS) and Centre National d’Etudes Spatiales (CNES).  HSL is grateful to the
hospitality of the Observatoire de Paris and La Clemence (1204 CH Geneva) during the 
initial drafting  of this manuscript.  We thank the referee for helpful comments
that served to clarify the discussion.

\end{acknowledgements}

\bibliographystyle{apj}

\end{document}